\documentclass{article}
\usepackage{spconf,amsmath,graphicx}
\usepackage{subfigure}
\usepackage{mathtools}

\DeclarePairedDelimiter\floor{\lfloor}{\rfloor}
\usepackage{subfig}
\title{SubSpectralNet -- USING SUB-SPECTROGRAM BASED CONVOLUTIONAL NEURAL NETWORKS FOR ACOUSTIC SCENE CLASSIFICATION}
\name{Sai Samarth R Phaye$^{1}$, Emmanouil Benetos$^{2,3}$, and Ye Wang$^1$\thanks{EB is supported by RAEng Research Fellowship RF/128 and EPSRC Grant EP/R01891X/1. This work was supported by The Alan Turing Institute under the EPSRC grant EP/N510129/1.}}
\address{$^1$School of Computing,  National University of Singapore, Singapore \\$^2$School of EECS, Queen Mary University of London, UK\ \ \ $^3$The Alan Turing Institute, UK
}

\usepackage{graphicx}
\usepackage{booktabs}
\usepackage{multirow}
\ninept

\begin{document}

\maketitle
\begin{abstract}
Acoustic Scene Classification~(ASC) is one of the core research problems in the field of Computational Sound Scene Analysis. In this work, we present SubSpectralNet, a novel model which captures discriminative features by incorporating frequency band-level differences to model soundscapes. Using mel-spectrograms, we propose the idea of using band-wise crops of the input time-frequency representations and train a convolutional neural network~(CNN) on the same. We also propose a modification in the training method for more efficient learning of the CNN models. We first give a motivation for using sub-spectrograms by giving intuitive and statistical analyses and finally we develop a sub-spectrogram based CNN architecture for ASC. The system is evaluated on the public ASC development dataset provided for the ``Detection and Classification of Acoustic Scenes and Events''~(DCASE) 2018 Challenge. Our best model achieves an improvement of +14\% in terms of classification accuracy with respect to the DCASE 2018 baseline system. Code and figures are available at \textit{https://github.com/ssrp/SubSpectralNet}
\end{abstract}
\section{INTRODUCTION}
\label{sec:intro}

The problem of recognizing the acoustic soundscapes and identifying the environment in which a sound is recorded is known as Acoustic Scene Classification \cite{book, eronen}. The objective is to assign a semantic label (acoustic scene) to the input audio stream that characterizes the type of environment in which it is recorded -- for example shopping mall, airport, street. The problem has been very well explored as a single-label classification task \cite{dcase2013, dcase2016}. Due to the possible presence of diverse sound events in a sound scene, developing a descriptive representation for ASC is known to be a difficult task \cite{roger2018acm}.

DCASE Challenges, started in 2013, provide benchmark data for computational sound scene analysis research, including tasks for detection and classification of acoustic scenes and events, motivating researchers to further work in this area. Looking at the current trend of challenge submissions in the ASC task, it is clear that researchers are moving towards using deep learning methods for system development \cite{dcase2013, dcase2016,dcase2017}. This is because of the fact that the current hand-crafted methods are not sufficient to capture the discerning properties of soundscapes \cite{lagrange2015bag}. With time, data-driven approaches are taking over conventional methods which involve more expert knowledge for designing and choosing features. Most published systems typically use a combination of audio descriptors and learning techniques, with a growing inclination towards deep learning \cite{bisot2017,treelabel}.%Data-driven methods help to learn discriminative features which captures the inter- and intra-class variances in the data.

The literature of ASC research is vast and a lot has been done in system design. Earliest works in this field have tried to use numerous methods from speech recognition (for example, using features like Mel-frequency cepstral coefficients \cite{mfcc_1}, normalized spectral features, and low-level features \cite{eronen,lowlevel_2} like the  zero-crossing rate). General architecture follows a pipeline based on extracting frame-by-frame hand-crafted audio features or learning them using various methods like matrix decomposition of spectral representations (log mel-spectrograms \cite{nmf}, Constant-Q transformed spectrograms \cite{cqt}), and then performing machine learning based classification. The final decision is a combination of frame wise outputs, for example, by using majority voting or mean probability. Many systems incorporate deep learning approaches, generally by using some kind of time-frequency representation as the input and training deep neural networks~(DNNs) or CNNs \cite{cnnfirstASC, piczak2015environmental}. Some methods also exploited ideas from the image processing literature, for example, training a classifier using the histogram of gradient representations over spectrograms of audio frames \cite{hog_1, hog_2}.

CNNs are extensively used in ASC. Some systems incorporate the use of convolutional layers with large receptive fields~(kernels) to capture global correlations in the spectrograms \cite{bigKernel1, bigKernel2}, while some use smaller kernels focusing on local spatial data %
\cite{smallKernel1, cnnfirstASC}. Rather than aiming for state of the art results, our goal is to show how sub-spectrograms could be used in CNNs to infer sound scenes more efficiently. Our work shows that depending on the scene class, there is a specific frequency band showing most activity, hence providing discriminative features for that class; to the authors' knowledge this has not been considered in earlier studies. We first develop a motivation for using spectrogram crops, which we term \emph{Sub-spectrograms}. Finally, we propose a CNN model, \emph{SubSpectralNet}, to make use of the Sub-spectrograms to capture more enhanced features, hence resulting in superior performance over a model with similar parameters which does not incorporate sub-spectrograms (discussed in Section \ref{sec:experiments}). For all experiments, we used the DCASE 2018 ASC development dataset \cite{dcase2018baseline} having 6122 two-channel 10-second samples for training and 2518 samples for testing, divided into ten acoustic scenes.

The rest of the paper is divided as follows -- in Section \ref{sec:motivation}, we develop a basic statistical model for ASC which we use as the motivation to design the proposed CNN architecture. Section \ref{sec:methodology} discusses the methodology used to develop the CNN model and Section \ref{sec:experiments} describes various experiments performed to prove the efficacy of the system. Finally, we conclude the work in Section \ref{sec:conclusion}.

% contributions section
\section{STATISTICAL ANALYSIS OF SPECTROGRAMS}
\label{sec:motivation}

Magnitude spectrograms are two-dimensional representations over time and frequency, which are very different from real life images. In spectrograms, there is a clear variation in the frequency axis. While images have local relationships over both spatial dimensions, spectrograms have definitive local relationships in the time dimension, but not in the frequency dimension. In the frequency dimension, for some types of sounds there are local relationships (e.g. sounds that have broadband spectra like noise-like sounds), sometimes they have non-local relationships (e.g. harmonic sounds, where there are relationships between non-adjacent frequency bins), and sometimes there are simply no local relationships at all.

\begin{figure}
\centering
\subfigure[]{\includegraphics[width=0.4\columnwidth]{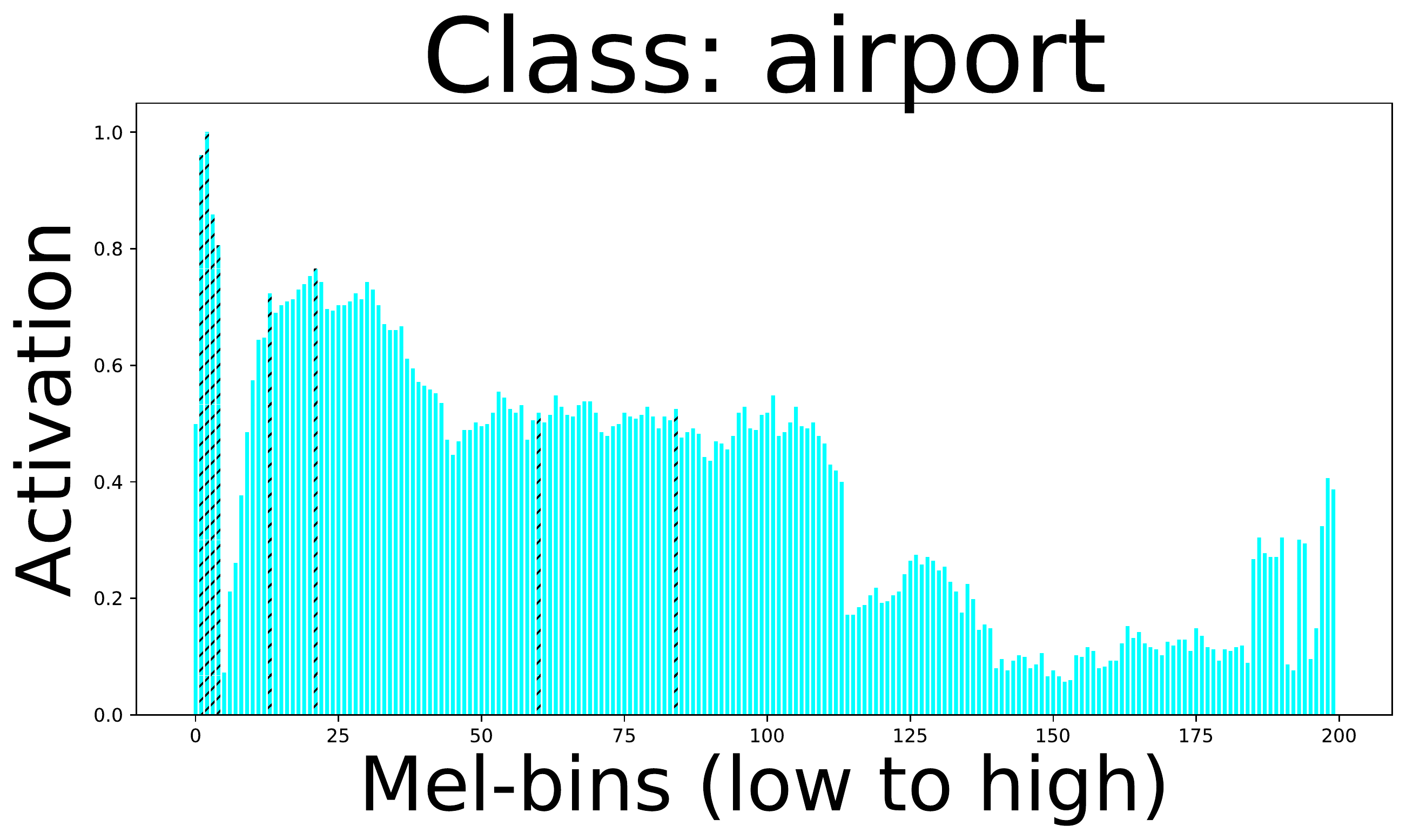}}
% \subfigure[]{\includegraphics[width=0.4\columnwidth]{ActivationsCombined-3.pdf}}
\subfigure[]{\includegraphics[width=0.4\columnwidth]{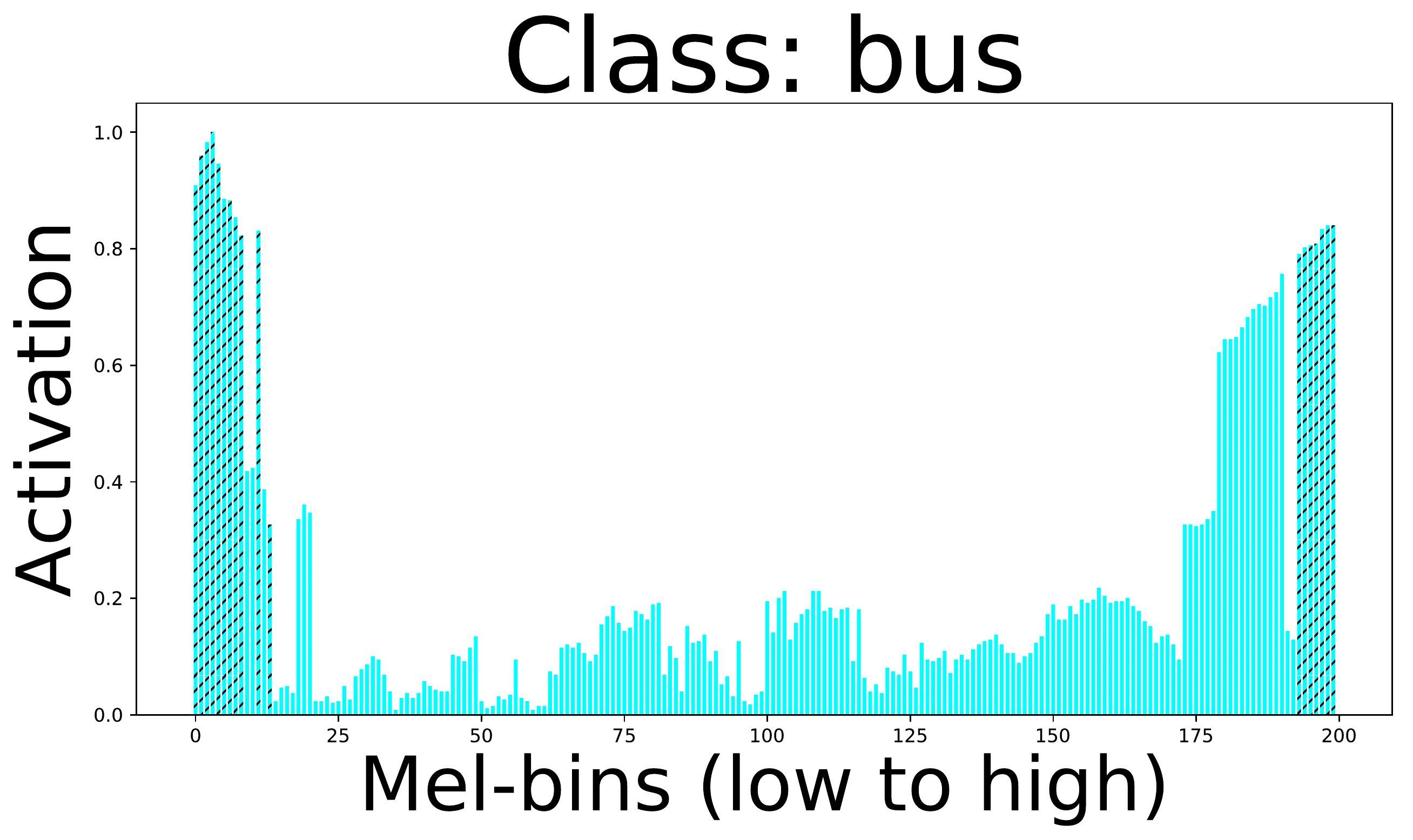}}
\subfigure[]{\includegraphics[width=0.4\columnwidth]{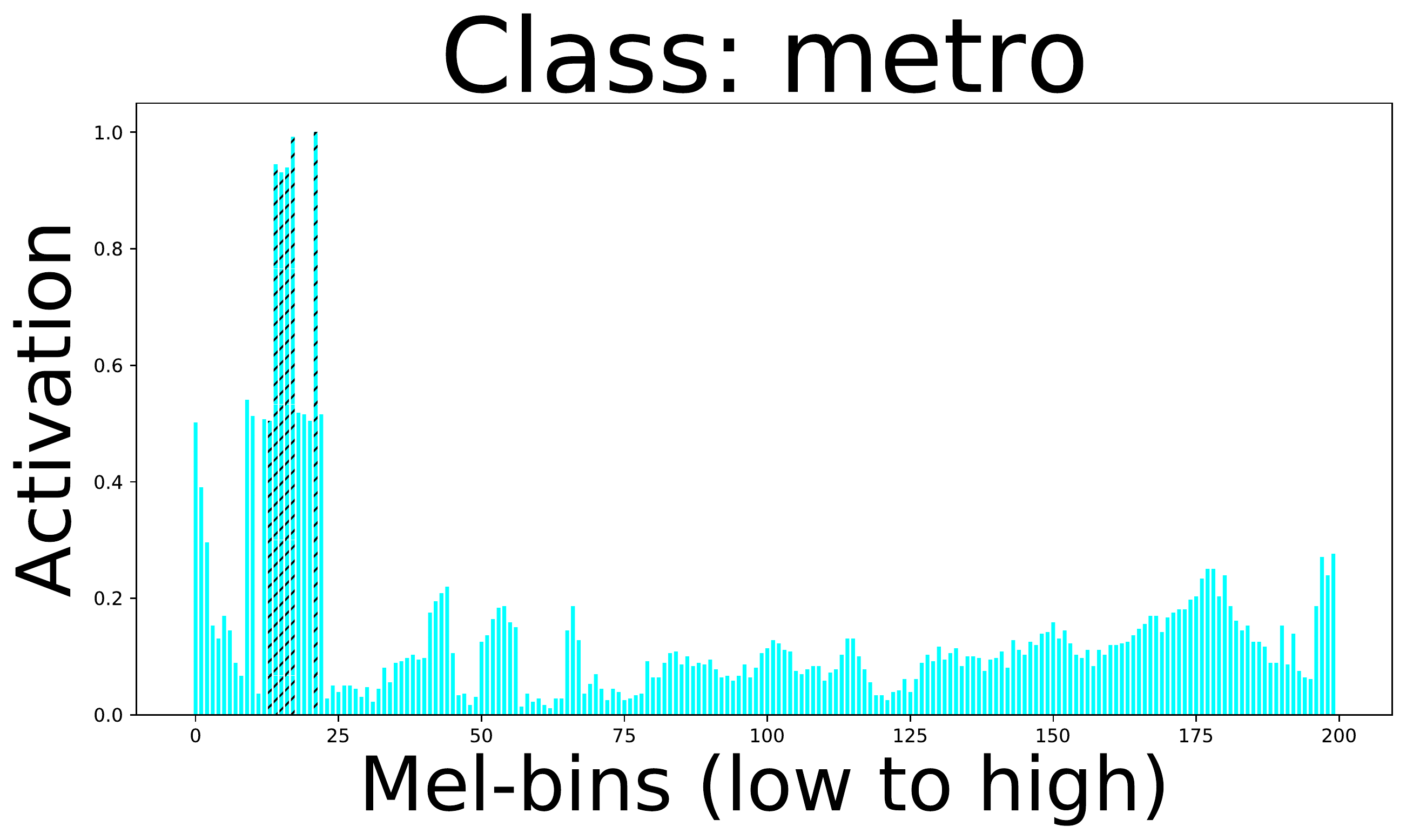}}
\subfigure[]{\includegraphics[width=0.4\columnwidth]{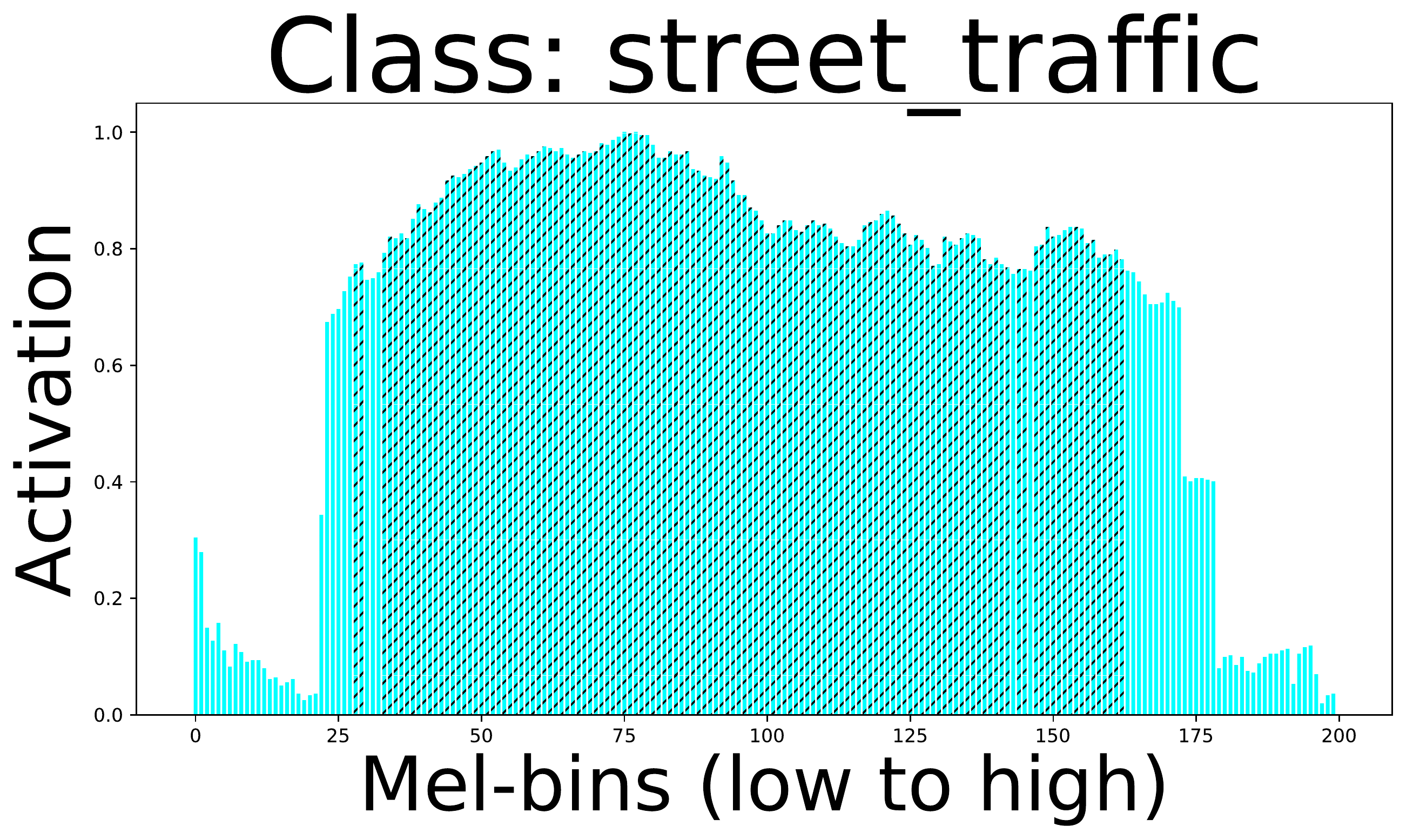}}
\subfigure[]{\includegraphics[width=0.4\columnwidth]{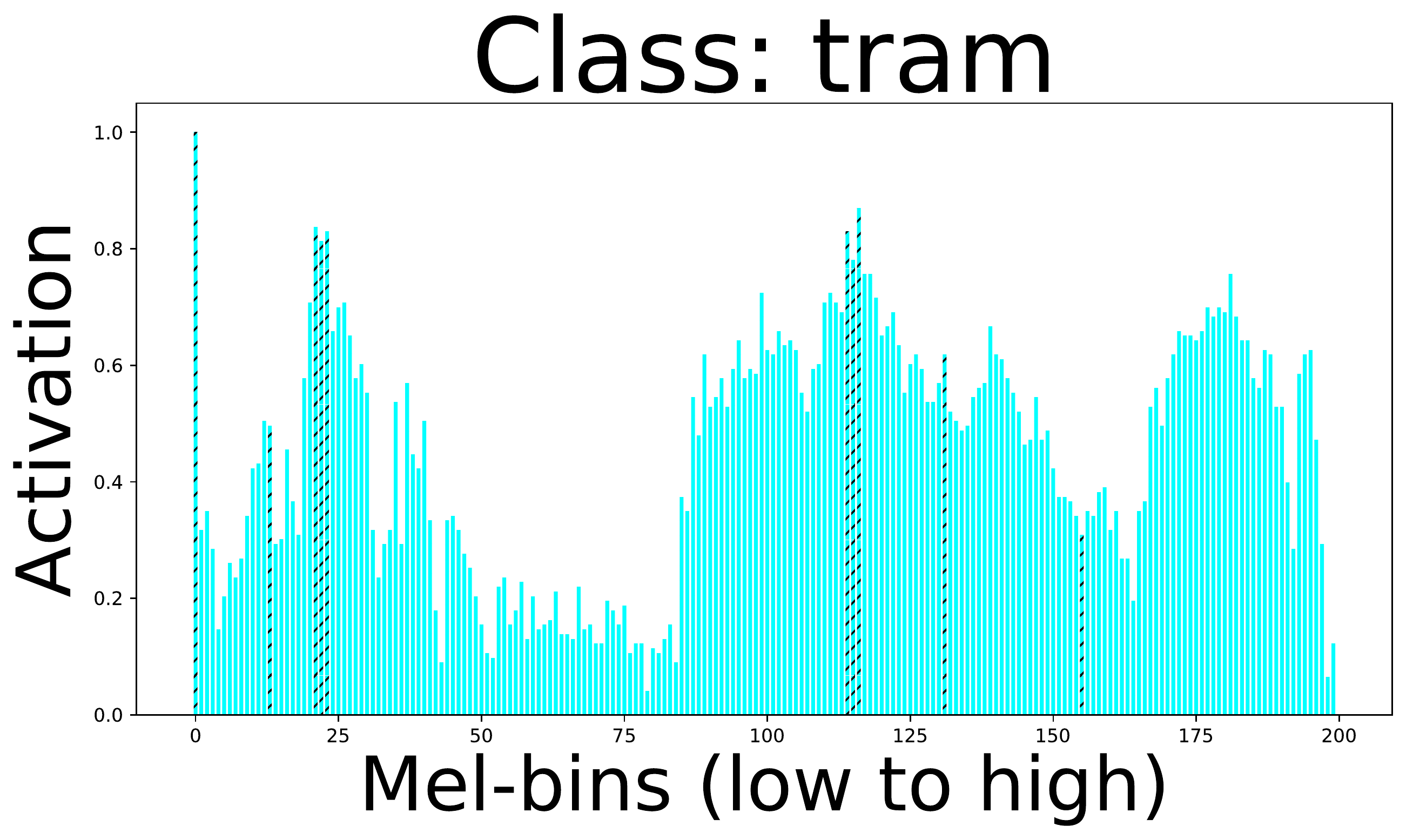}}
% \subfigure[]{\includegraphics[width=0.4\columnwidth]{ActivationsCombined-5.pdf}}
\subfigure[]{\includegraphics[width=0.4\columnwidth]{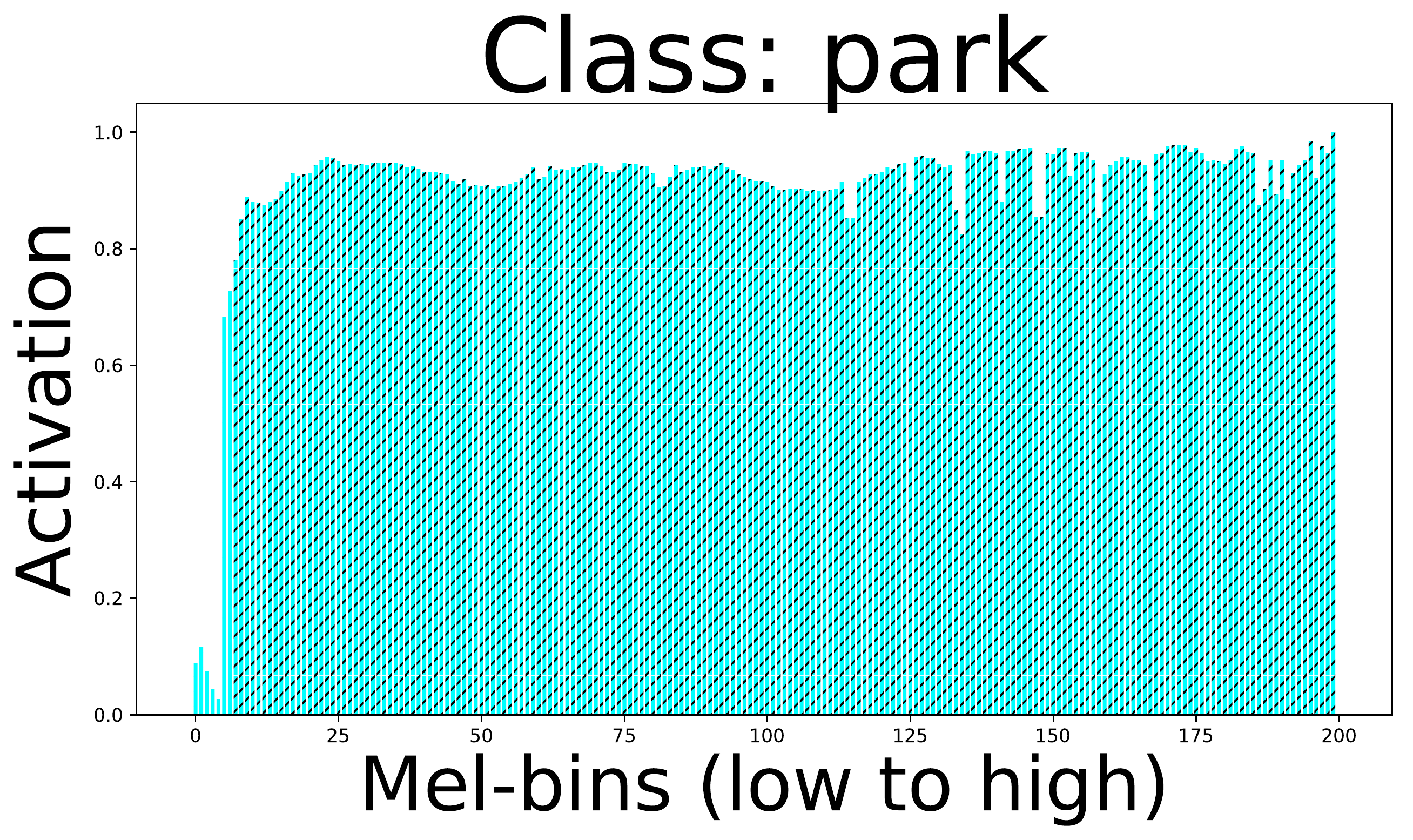}}
% \subfigure[]{\includegraphics[width=0.45\columnwidth]{ActivationsCombined-7.pdf}}
%\subfigure[]{\includegraphics[width=0.4\columnwidth]{ActivationsCombined-6.pdf}}

\caption{Histogram of activation of mel-bins for some sound scene classes. We can infer the importance of specific mel-bins for specific classes from these histograms. This is also intuitively true, for example, in an airport or in a metro, audio may have dominant and discriminative low-frequency noise, and lower bands of the spectrograms show more activation for these classes.}
\label{fig:histograms}
\end{figure}

We first create a simple mathematical model to gain more insights on how CNNs could leverage time-frequency features efficiently. We extract log mel-spectrograms using a 2048-point short time Fourier transform~(STFT) on 40ms Hamming windowed frames with 20ms overlap and then transform this into 200 Mel-scale band energies. Finally, the log of these energies is taken. Next, we perform bin-wise normalization of the sample space and obtain 6122 samples having $200 \times 500$ (mel-bins $\times$ time-index) feature size. Now, we concatenate all the samples of the same class in the time dimension and take the average of the temporal direction to obtain ten distributions having 200 vector-size, one for each class. We observe that there is a clear variation in the class-wise activation of different mel-bins. For more clarity, we perform bin-wise classification of test samples using the ten 200D reference mean vectors.

\begin{figure}[t]
  \centering
  \includegraphics[width=0.8\columnwidth]{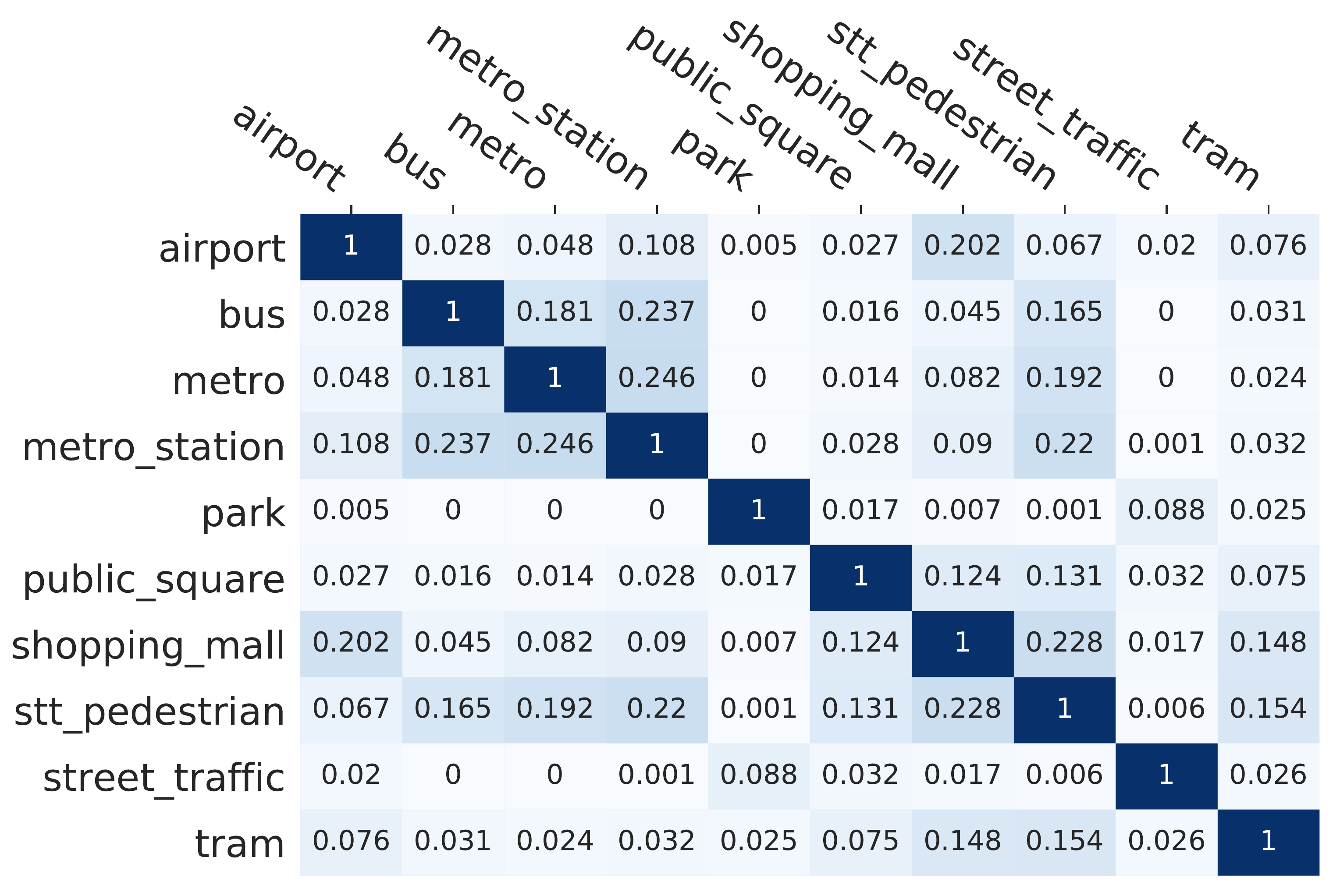}
  \caption{Resultant Chi-Square Distance Matrix.}
   \label{fig:ChiSquare}
\end{figure}

For each test sample, we compute the mean of temporal direction to get a 200D vector and the hypothesis is that this vector should have a similar distribution as that for the mean vector of the corresponding class. Mathematically, we compute the L$^2$ distance with the reference mean vectors and the class for which this distance has the minimum value should be the correct label.

Now, instead of computing the distance measure for the whole 200D vector, we compute separate distances for each mel-bin because we are interested in analyzing how those bins are activated for specific classes. This is equivalent to saying that we have 200 small classifiers. Finally, using these 200 outputs for all the test samples, we create one normalized histogram for each class, in which we have frequencies of correct classifications of corresponding mel-bins, shown in Fig.~\ref{fig:histograms}. We also calculate the chi-square distance between these histograms to see how similar class-wise distributions are. For this, we normalize the histograms with maximum value to one, and then compute the distance. The lesser the distance, the more confusion exists between those classes. We aim to obtain a matrix which has some resemblance with a confusion matrix. For that, after getting the $10 \times 10$ symmetrical matrix having distances between the classes, we normalize the matrix by dividing with the maximum value. Then, we apply the following mathematical transform:
\begin{gather}
 x^{new}_{ij} = 1 - e^{-kx_{ij}}
\end{gather}
where $x_{ij}$ is the prior distance value and $i, j$ are the matrix indices. $k$ is a constant parameter which when increased, enhances the differences of values on the higher range. We used $k$ as $10$ so that the matrix resembles a confusion matrix. Next, we normalize again the matrix by dividing with the maximum value and lastly, subtract these values from one. The output of this is shown in Fig.~\ref{fig:ChiSquare}. We also compute the Kullback-Leibler divergence \cite{kldivergence} and Hellinger distance \cite{hellinger} over these histograms and they result in a very similar matrix, which shows that the statistical model is robust. We can clearly see that some classes are having higher confusions (for example, \textit{``metro\_station"} and \textit{``metro"}; \textit{``shopping\_mall"} and \textit{``airport"}), which resembles the confusion matrices obtained from the baseline model results \cite{dcase2018baseline} and proposed CNN model (shown in Figure \ref{fig:confusionMatrices}).

In the histograms obtained, we observe a definite variation of activation of mel-bins and sub-bands, which is specific to every scene. For example, the \textit{``metro"} class has more activation in lower frequency bins; the \textit{``bus"} has less activation in mid frequency bins. For \textit{``park"} or \textit{``street\_traffic"}, nearly all mel-bins are active and from the DCASE 2018 baseline result \cite{dcase2018baseline}, we can see that these classes have relatively superior performance. We use these observations to develop SubSpectralNet, which is discussed in the next section. 

% 'airport','bus','metro','metro_station','park','public_square', 'shopping_mall','street_pedestrian','street_traffic','tram'

\section{DESIGNING SUBSPECTRALNET}
\label{sec:methodology}

We start off with the DCASE 2018 baseline system for the ASC task and gradually develop the proposed network. The baseline system is based on a CNN, where mel-band energies with 40 mel-bins are extracted for every sample with 40 millisecond frame size and 50\% overlap using 2048-point STFT. The samples are further normalized and the size of each sample is $40 \times 500$. These samples are passed to a CNN consisting of two layers with \textit{same} padding in order -- 32 kernels and 64 kernels, each having kernels of $7 \times 7$ size, batch normalization and ReLU activation. After each conv-layer, a max-pooling layer of $5 \times 5$ and $4 \times 100$ pool-size respectively is used to decrease the size of the feature space and a dropout rate of 30\% is applied to prevent over-fitting. Finally, a fully connected~(FC) layer with 100 neurons is used over the flattened output, which is further connected to the output~(softmax) layer.

We train DCASE 2018 baseline models on different channels of the audio dataset -- left channel, right channel, average-mono channel and lastly, both channels to the CNN model. The best results are obtained using both channels which is expected as binaural input would give more information on the prominent sound events in soundscapes, for example, a car passing by in \textit{``street\_traffic"}.

\subsection{Incorporating Sub-spectrograms}
From the analysis in Section \ref{sec:motivation}, we infer that using bigger convolutional kernels over spectrograms is not a good idea because it tends to combine global context and we lose the local time-frequency information. We perform an experiment (discussed in Section \ref{sec:experiments}) in which we gradually increase the size of the kernels in the first conv-layer of the baseline system. The accuracy decreases with the increase in kernel size. Spectrograms have a definite variation in the frequency dimension. Using smaller convolutional kernels over complete spectrograms works fine because CNNs are very powerful in fitting these receptive fields to understand the variances in the data. But the fact that spectrograms have these variations could be advantageous.

\begin{figure}[t]
  \centering
  \includegraphics[width=\columnwidth]{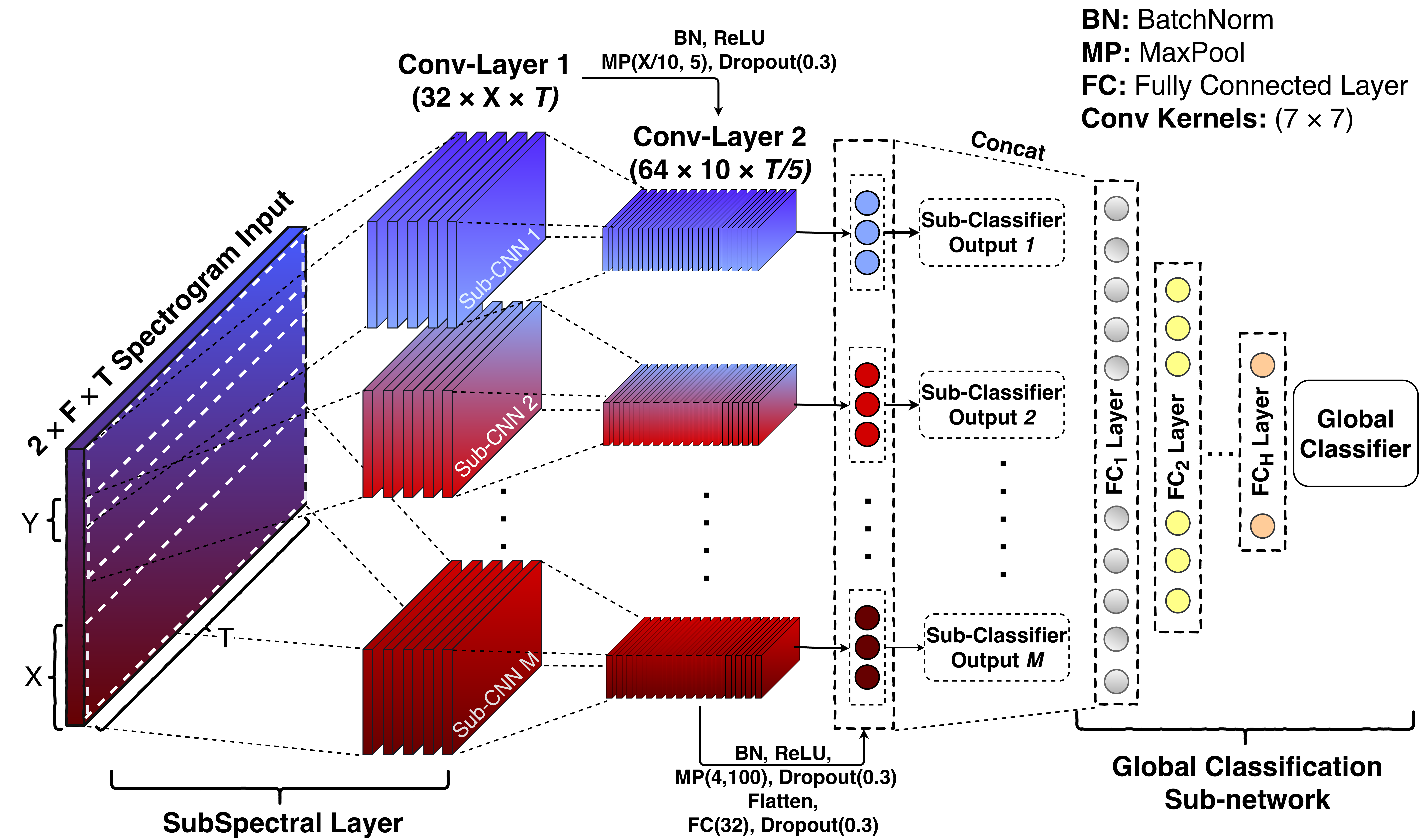}
  \caption{Proposed pipeline of SubSpectralNet.}
   \label{fig:subspectralnet}
\end{figure}

Building upon this idea, we propose SubSpectralNet and its architecture is shown in Figure~\ref{fig:subspectralnet}. SubSpectralNet essentially creates horizontal slices of the spectrogram and trains separate CNNs on these sub-spectrograms, finally acquiring band-level relations in the spectrograms to classify the input using diversified information.

We extract the log mel-energy spectrograms for the $N$ samples and perform bin-wise normalization. For creating sub-spectrograms, we design a new %Keras % EB: this is an implementation aspect, which can be discussed in Section 4.
layer (we term it as \emph{SubSpectral Layer}) which splits the spectrogram into various horizontal crops. It takes three inputs -- input spectrogram ($C \times F \times T$ dimension, $C$, $F$ and $T$ being number of channels, mel-bins and time-indices respectively), sub-spectrogram size $X$ and mel-bin hop-size~(vertical hop) $Y$. This results in $M$ frequency-time sub-spectrograms of $C \times X \times T$ dimension for every sample, where $M = \floor*{1 + (F-X)/Y}$.

\subsection{SubSpectralNet -- Architecture Details}
Two-channel sub-spectrograms are independently connected to 2 conv-layers with \textit{same} padding and kernel-size of $(7, 7)$ having 32 and 64 kernels respectively. After each conv-layer, there is a batch normalization layer, ReLU activation layer,  max-pooling layers of $(X/10, 5)$ and $(4, 100)$ size respectively, and finally a dropout of 30\%. After the second pooling, we flatten the layer and add an FC layer with 32 neurons with ReLU activation and 30\% dropout, followed by the softmax layer. We call these sub-classifiers of the SubSpectralNet. We do not remove these softmax outputs from the final network because this enforces them to learn to classify the sample based on only a part of spectrogram. We keep most parameters same as the DCASE 2018 baseline model for fair comparison.  We believe that sub-spectrograms could be incorporated into more complex architectures \cite{separateoutputs, senets, densenets} that could be used to surpass the state of the art in ASC performance.

To capture the global correlation (or de-correlation) between frequency bands, we concatenate the FC (ReLU) layer of the sub-networks and train a DNN with $H$ hidden layers with $R_i$ neurons, where: 
$H = max(\floor*{log_2(M)}-1, 0); R_i = 2^{6 + H - i}, 1 \leq i \leq H$.
We term this as the \emph{global classification sub-network}. 

All cross-entropy errors from the global and sub-classifiers are back-propagated simultaneously to train a single network. The sub-classifiers learn to classify using specific bands of spectrograms, while the global classifier combines and learns discerning information at the global level. This modification of training method results in improved performance and faster convergence of the model with minimal addition to the complexity \cite{separateoutputs}.

\begin{figure*}
\centering
\subfigure[]{\includegraphics[width=0.5\columnwidth]{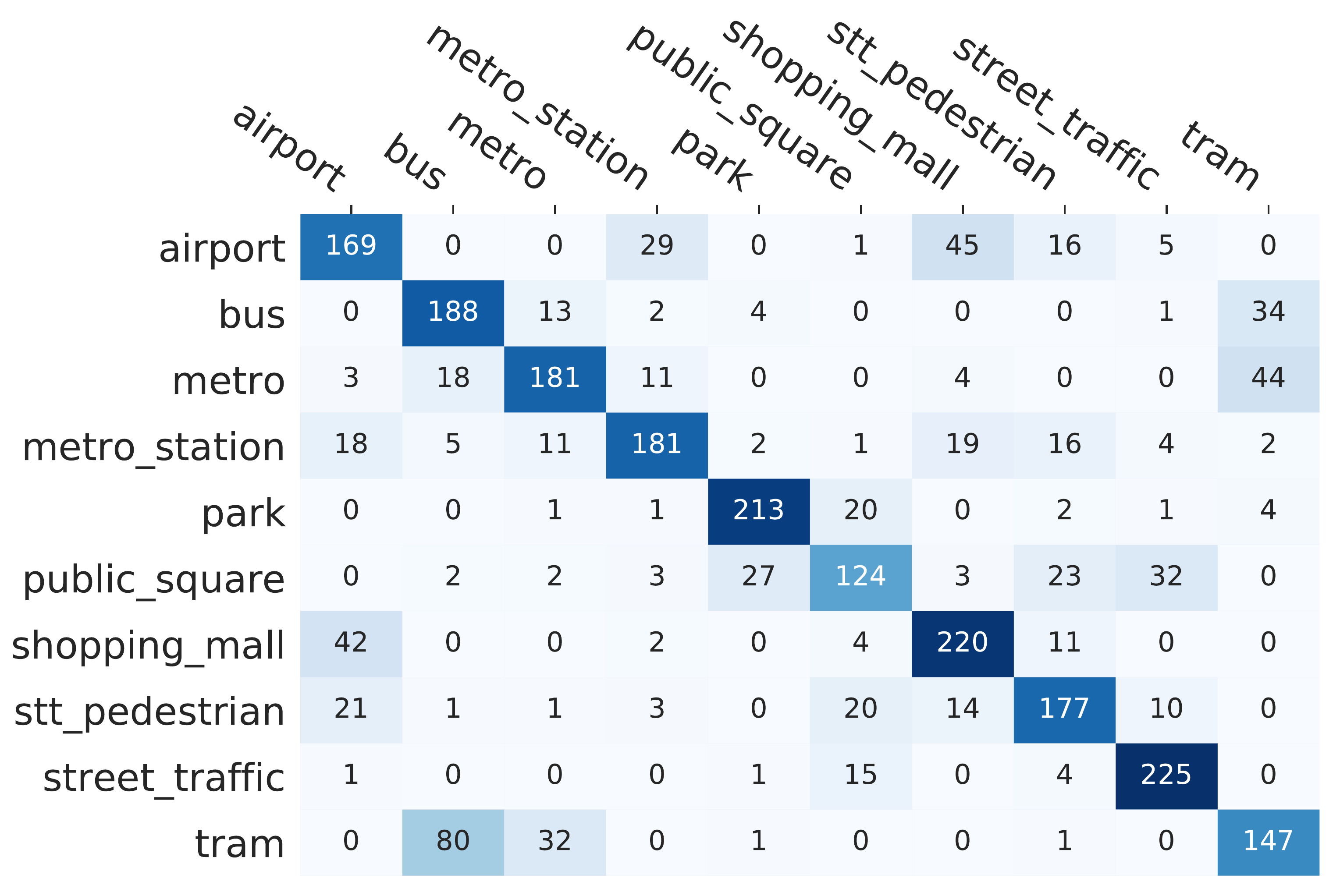}}
\subfigure[]{\includegraphics[width=0.5\columnwidth]{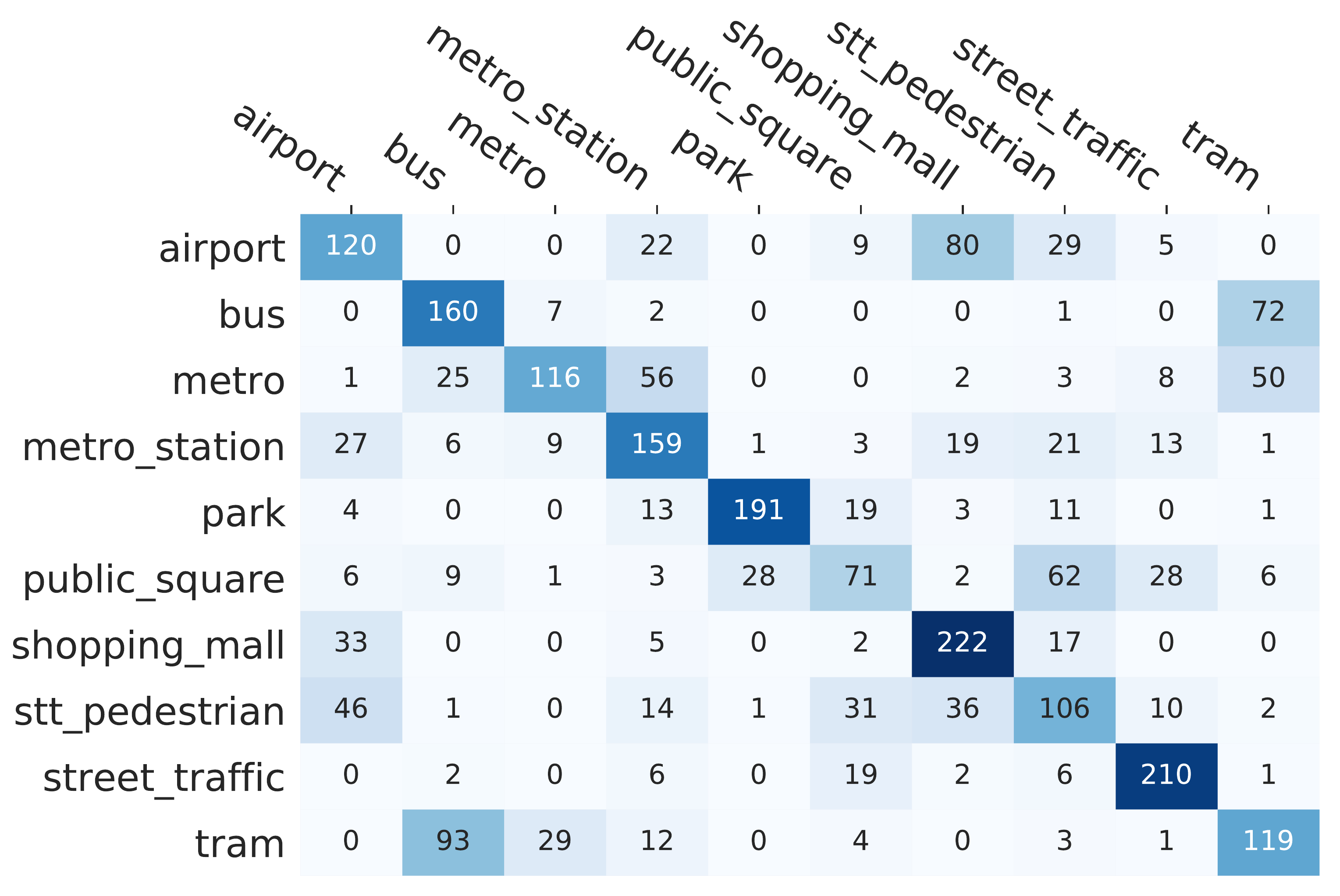}}
\subfigure[]{\includegraphics[width=0.5\columnwidth]{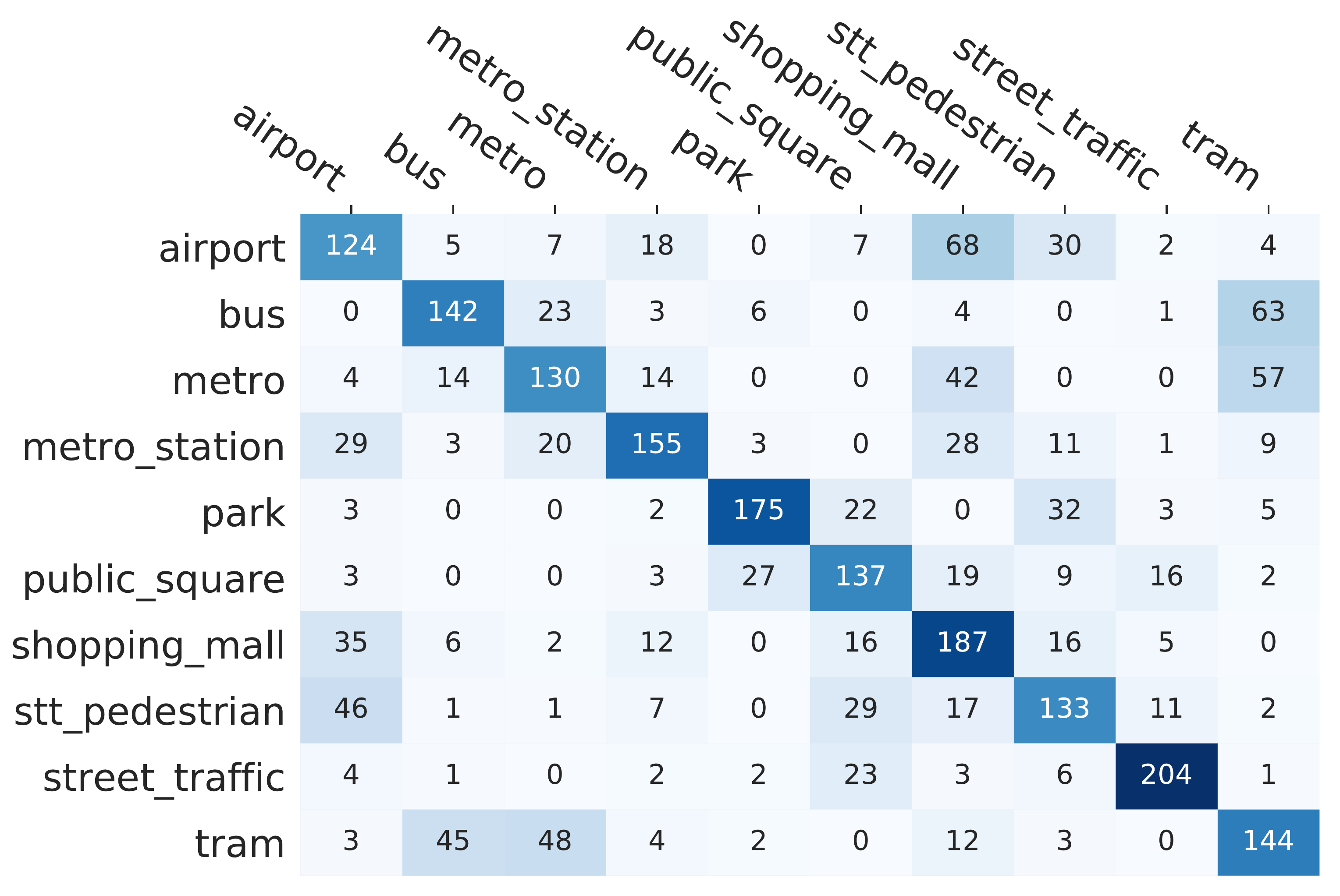}}
\subfigure[]{\includegraphics[width=0.5\columnwidth]{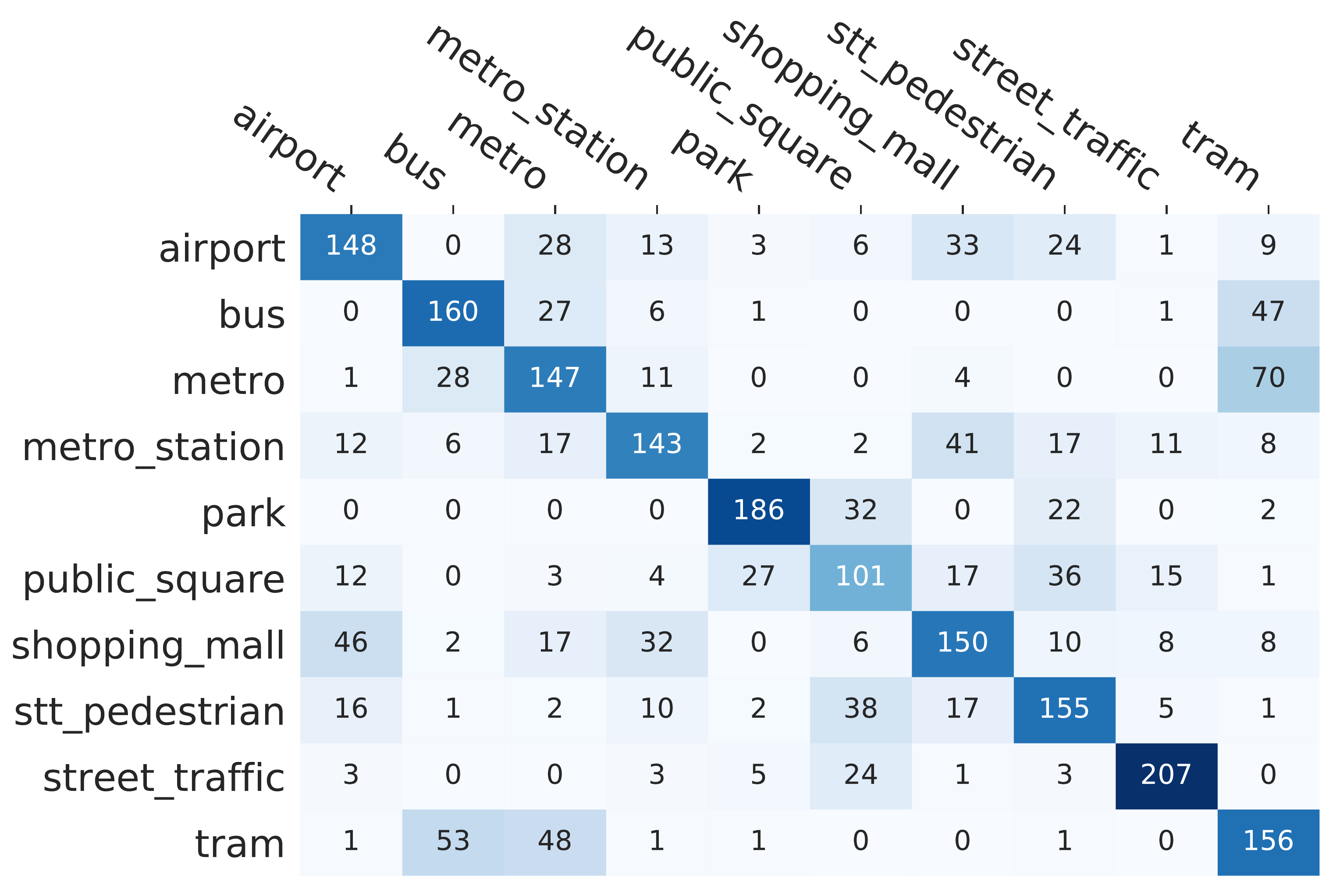}}
\caption{Confusion matrices obtained from a SubSpectralNet trained over 40 mel-bin spectrograms, 20 sub-spectrogram size, 10 mel-bin hop-size, hence 3 sub-classifiers and one global classifier. Matrices are obtained for (a) Global Classifier, (b) High-frequency Band Sub-Classifier (21-40 mel-bins), (c) Mid-frequency Band Sub-Classifier (11-30 mel-bins) and (d) Low-frequency Band Sub-Classifier (1-20 mel-bins).}
\label{fig:confusionMatrices}
\end{figure*}

We create confusion matrices (shown in Fig.~\ref{fig:confusionMatrices}) from the output of these sub-classifiers and the global classification model discussed in Section~\ref{sec:experiments}. We observe that the statistical motivation given in Section \ref{sec:motivation} fits well with the results. For example, for the \textit{``airport"} class, statistical distribution says that lower frequencies are more effective in classification. The same is shown in the confusion matrix where the low-band sub-classifier shows better results for this class. For the \textit{``bus"} class, the mid-band sub-classifier shows relatively better results. For most classes, the global classifier achieves better results than any sub-classifier. It is interesting to note that for some classes like \textit{``public\_square"} or \textit{``tram"}, a sub-classifier performs better than the global classifier, which could mean that using the complete spectrogram adds outliers and it is better to use a specific band of spectrogram in such case.

% Talk about the basic sub-spectrogram model.
% 1. Separate training, ensemble
% 2. Global single training
% 3. Global training and separate training parallel

% talk about design considerations: why separate back propagation. 

% add separate classifier accuracy plot 40.20.10

\section{EXPERIMENTS}
\label{sec:experiments}

We demonstrate the potential of SubSpectralNet on the DCASE 2018 development public dataset (Task 1A) and compare the results with DCASE 2018 baseline. We use \textit{dcase\_util} toolbox \cite{dcase_util} to the extract features from the DCASE 2018 dataset. We implement SubSpectralNet in Keras with TensorFlow backend and experiments are performed on an NVIDIA Titan Xp GPU having 12GB RAM. We train all models 3 times for 200 epochs and report the average-best accuracy. The learning rate is set to $0.001$ with Adam as the optimizer. Following are the experiments we perform in this work:

\begin{figure}[t]
  \centering
  \includegraphics[width=0.65\columnwidth]{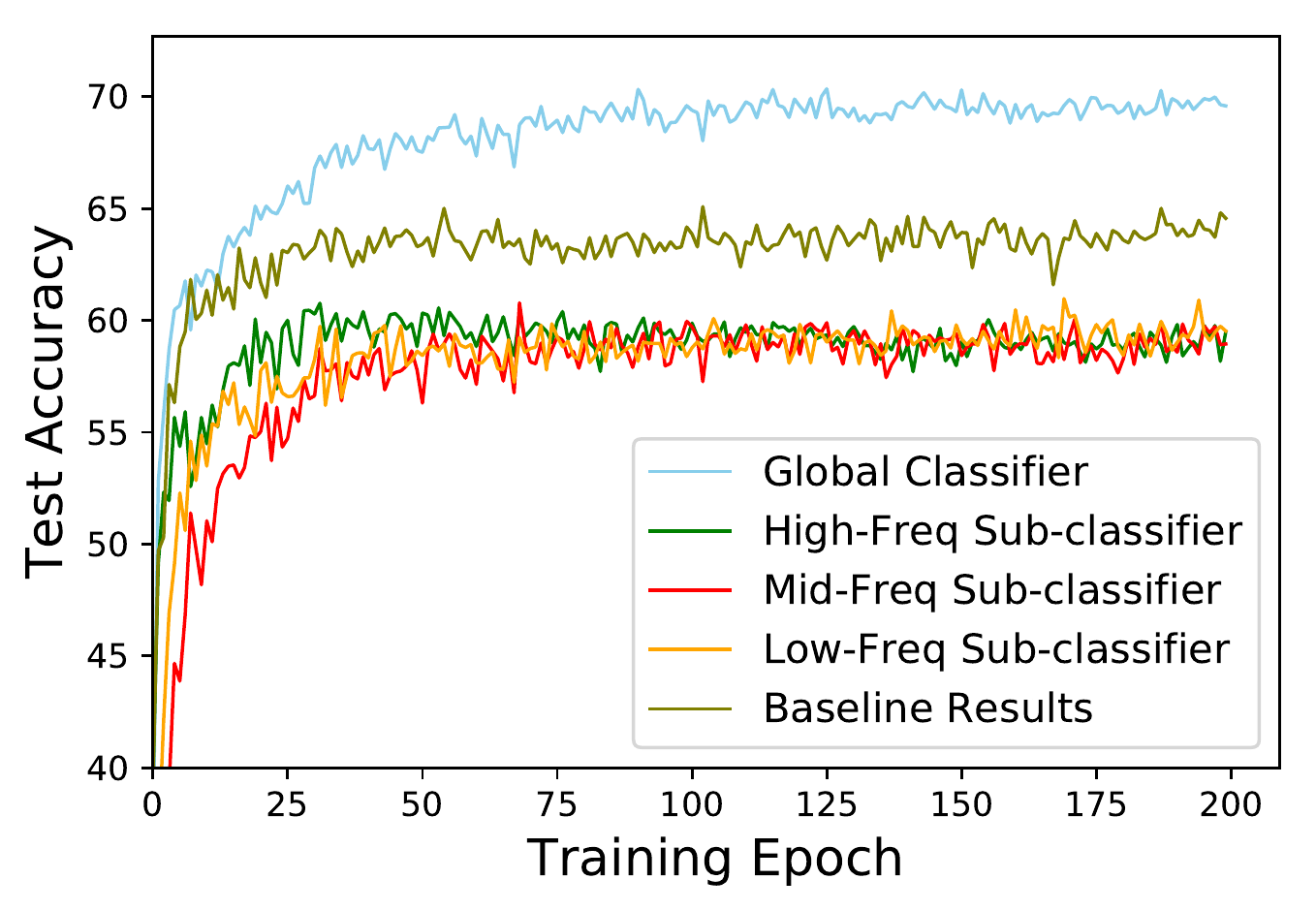}
  \caption{Comparison of performance between the multi-channel DCASE 2018 baseline model and SubSpectralNet on 40 mel-bin spectrograms, 20 sub-spectrogram size and 10 mel-bin hop-size.}
   \label{fig:accplot}
\end{figure}

We train DCASE 2018 baseline models on different channels of audio dataset and the test accuracy achieved are 63.24\% (left channel), 61.83\% (right channel), 64.91\% (average-mono channel) and \textbf{65.66\%} (stereo channels). We also train the DCASE 2018 baseline model for various kernel sizes of the first CNN layer -- $(7, 7)$, $(15, 15)$, $(25, 25)$ and $(35, 35)$. The corresponding test accuracies are \textbf{65.66\%}, 65.23\%, 65.08\% and 62.80\% respectively. This shows that bigger receptive fields tend to combine information on a bigger scale, hence losing local spatial information. As a result, we choose the kernel-size of $(7, 7)$ with stereo input for SubSpectralNet.

We train a SubSpectralNet model on 40 log mel-energy spectrograms with 20 sub-spectrogram size and 10 mel-bin hop-size (331K model parameters). The resultant accuracy we achieve from this is \textbf{72.18\%}. The confusion matrices computed for this model are shown in the Figure \ref{fig:confusionMatrices}. Also, we plot a curve of training epoch versus test accuracy, comparing the performance of the DCASE 2018 baseline (2-channel model) and this SubSpectralNet model, which is shown in Figure \ref{fig:accplot}. It can be seen that SubSpectralNet (global classifier) has a relatively faster convergence with superior test accuracy.

\begin{figure}
\centering
\subfigure[]{\includegraphics[width=0.31\columnwidth]{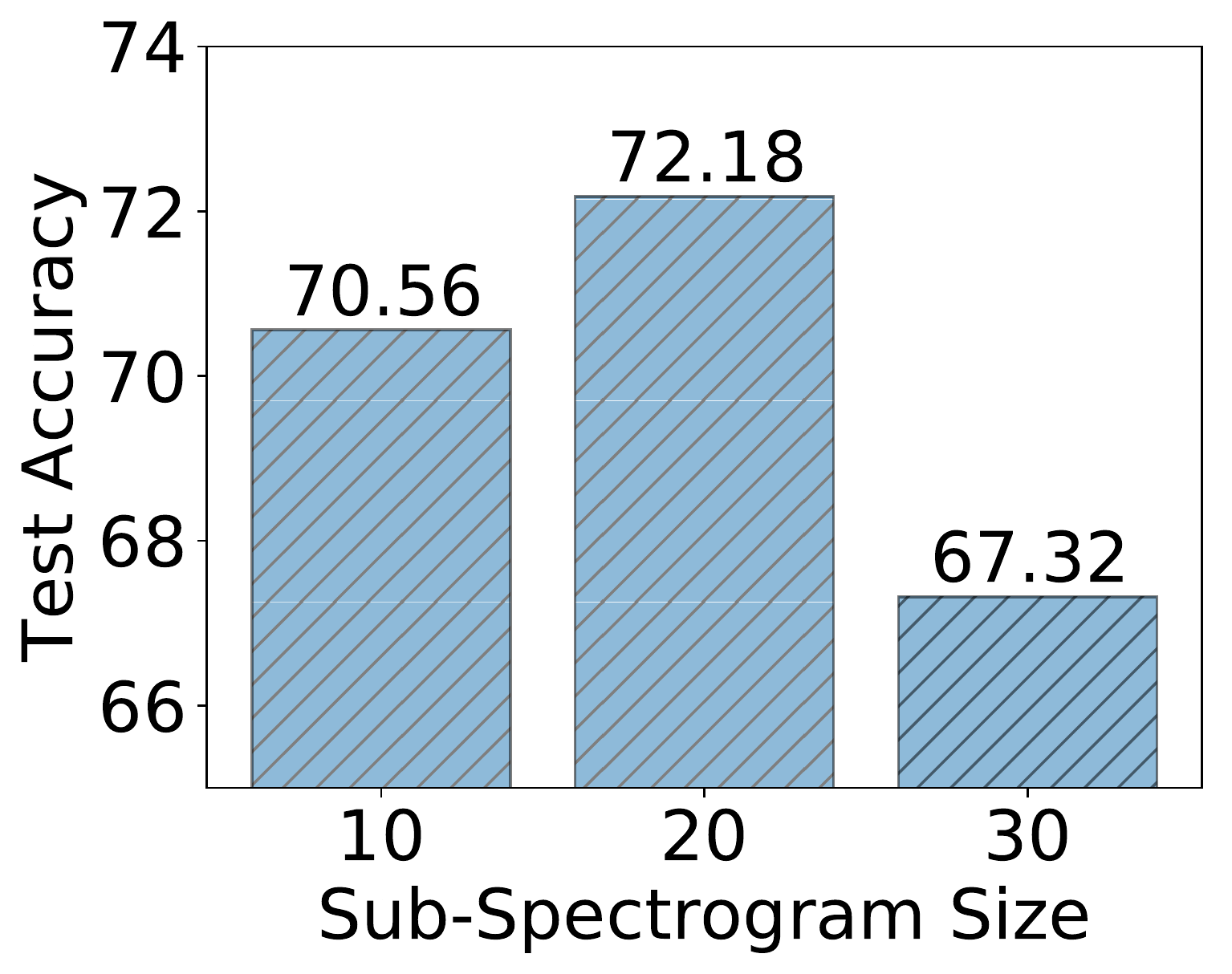}}
\subfigure[]{\includegraphics[width=0.31\columnwidth]{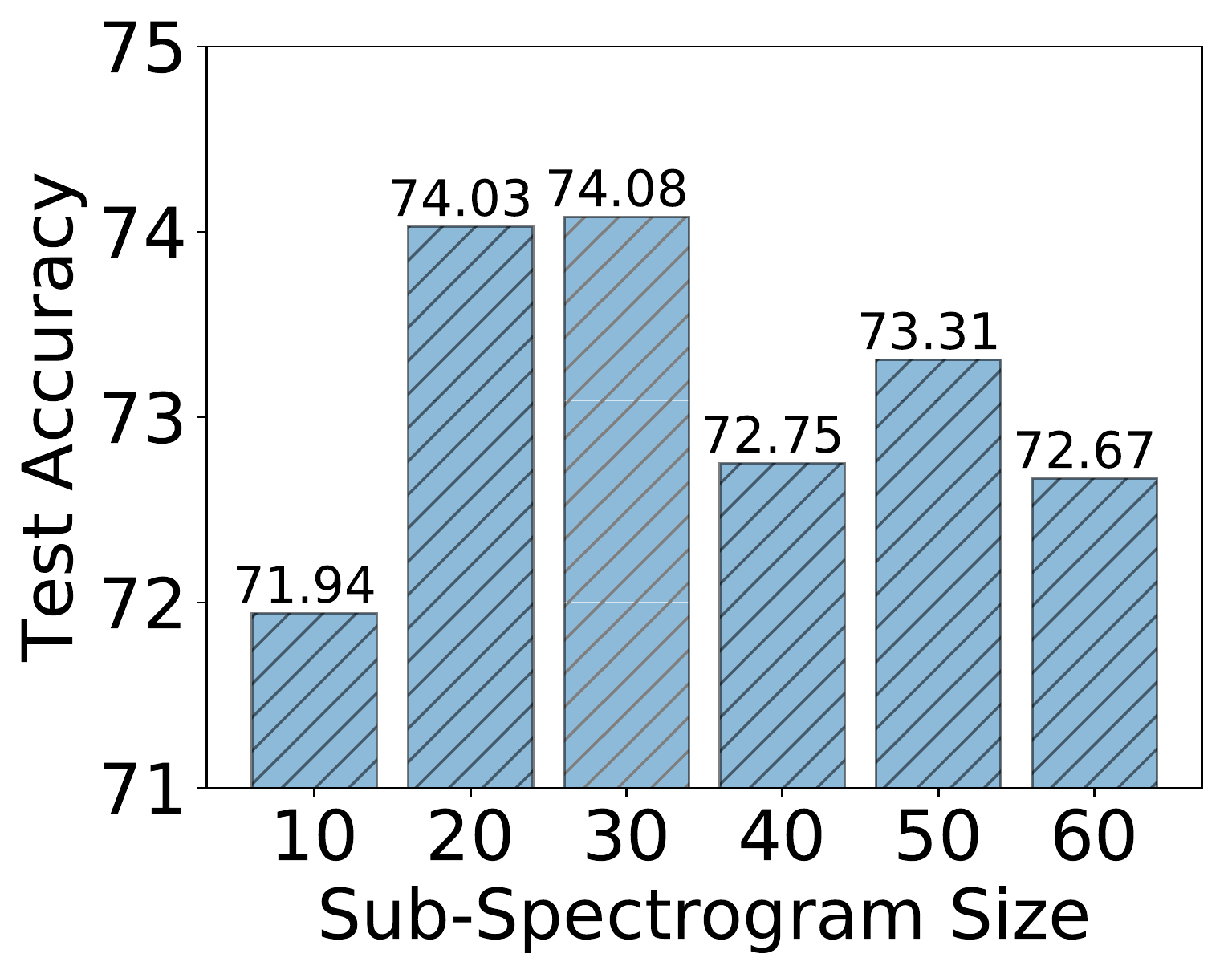}}
\subfigure[]{\includegraphics[width=0.34\columnwidth]{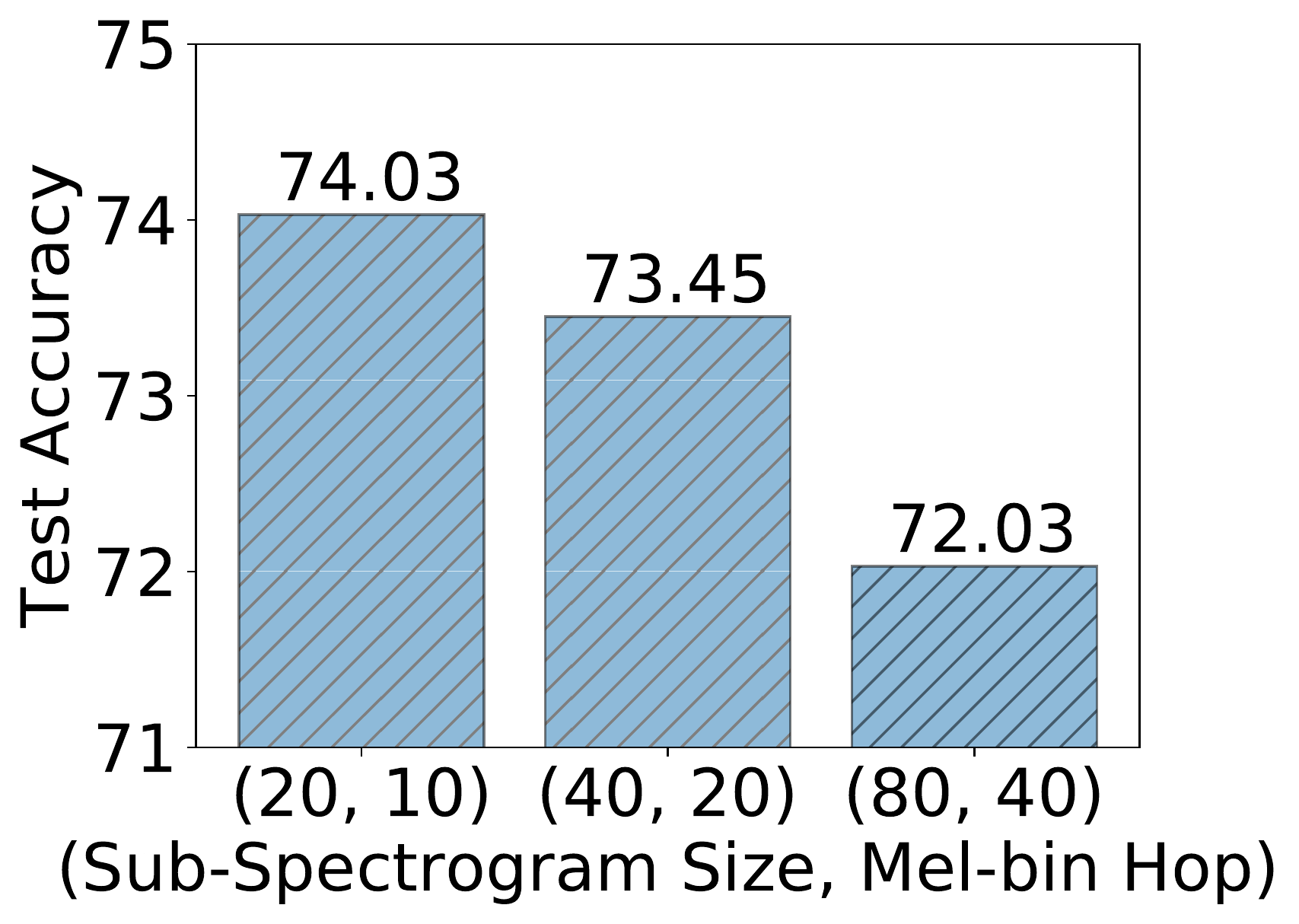}}

\caption{Results obtained by SubSpectralNet on -- (a) 40 mel-bin spectrogram and 10 mel-bin hop-size; (b) 200 mel-bin spectrogram with 10 mel-bin hop-size; (c) 200 mel-bin spectrogram, varying sub-spectrogram and mel-bin hop-size.}
\label{fig:ablations}
\end{figure}

To demonstrate the importance of sub-classifiers, we train a SubSpectralNet model excluding the Sub-Classifier (softmax layers) back-propagations and only use the global classification sub-network (330K model parameters). This achieves an accuracy of 68.79\%, comparing to 72.18\% with the sub-classifiers, which verifies the significance of the same. More experiments with 40 log mel-energy spectrograms are shown in Figure~\ref{fig:ablations} (a).

Parameters of a CNN are one of the major criteria to compare two models. The DCASE 2018 baseline model on 40 mel-bins has 117K parameters (using 2-channel input) and the SubSpectralNet model with 20 sub-spectrogram size and 10 mel-bin hop-size has 331K parameters. To prove the efficacy of SubSpectralNet, we modified the DCASE 2018 baseline model by doubling the number of kernels in both conv-layers (now 64 and 128). This model, having 434K parameters achieved 66.79\% accuracy which is \textbf{5.39\%} lower than the accuracy of proposed model. Hence, this justifies the fact that the idea of fitting separate kernels (training separate CNNs) over separate bands of spectrograms learns more salient features than directly training a CNN on spectrograms.

Considering that 200 mel-bin spectrograms achieve better performance than using lesser mel-bins \cite{bigKernel2}, we train a DCASE 2018 baseline model on 200 mel-bins and the accuracy achieved is 71.94\%. It is interesting to note that SubSpectralNet with 40 mel-bins can achieve comparably superior accuracy. We trained various SubSpectralNets on 200 mel-bins and the results are shown in Fig.~\ref{fig:ablations} (b) and (c). The best accuracy achieved was 74.08\%, by using 30 sub-spectrogram size and 10 mel-bin hop-size, which is an overall increase of \textbf{+14\%} over the DCASE 2018 baseline \cite{dcase2018baseline}.

% \begin{table}[t]
% \centering
% \caption{Comparison of various model variations after 50 epochs on MNIST dataset.}
% \label{param}
% \setlength{\tabcolsep}{8pt} % Default value: 6pt
% {\renewcommand{\arraystretch}{1.05}% for the vertical padding
% \resizebox{0.95\columnwidth}{!}{%
% \begin{tabular}{@{}|l|c|c|l|c|@{}}
% \toprule
% Model & Parameters & Sub-Spectrogram Size & Channels & Test Accuracy \\ \midrule
% \multirow{4}{*}{DCASE Baseline} & \multirow{3}{*}{116K} & \multirow{5}{*}{-} & Left & 63.24\% \\ \cmidrule(l){4-5} 
%  &  &  & Right & 61.83\% \\ \cmidrule(l){4-5} 
%  &  &  & Mono (avg) & 64.91\% \\ \cmidrule(lr){2-2} \cmidrule(l){4-5} 
%  & 117K &  & Both Channels & 65.66\% \\ \cmidrule(r){1-2} \cmidrule(l){4-5} 
% DCASE Bigger Network & 434K &  & \multirow{5}{*}{Both Channels} & 66.79\% \\ \cmidrule(r){1-3} \cmidrule(l){5-5} 
% \multirow{3}{*}{SubSpectralNet} & 441K & 10 &  & 70.56\% \\ \cmidrule(lr){2-3} \cmidrule(l){5-5} 
%  & 331K & 20 &  & \textbf{72.18\%} \\ \cmidrule(lr){2-3} \cmidrule(l){5-5} 
%  & 221K & 30 &  & 67.32\% \\ \cmidrule(r){1-3} \cmidrule(l){5-5} 
% \begin{tabular}[c]{@{}l@{}}SubSpectralNet\\ (without sub-classifiers)\end{tabular} & 330K & 20 &  & 68.79\% \\ \bottomrule
% \end{tabular}
% }
% }
% \end{table}

\section{CONCLUSIONS}
\label{sec:conclusion}
In this paper, we introduce a novel approach of using spectrograms in Convolutional Neural Networks in the context of acoustic scene classification. First, we show from the statistical analysis of Sec.~\ref{sec:motivation} that some specific bands of mel-spectrograms carry more discriminative information than other bands, which is specific to every soundscape. From the inferences taken by this, we propose SubSpectralNets in which we first design a new convolutional layer that splits the time-frequency features into sub-spectrograms, then merges the band-level features on a later stage for the global classification. The effectiveness of SubSpectralNet is demonstrated by a relative improvement of +14\% accuracy over the DCASE 2018 baseline model.

%Furthermore, addition of sub-classifier back-propagations helps to learn better feature maps on a band level, in turn forming enhanced features for the global classification. 

%The DCASE 2018 ASC dataset is the largest real-life and complex dataset on acoustic scenes so far, therefore performance of a network on this dataset could be treated as a substandard, compared to simpler datasets like DCASE 2013 ASC dataset. 

%In this work, we tried to leverage the properties of spectrograms when used over CNNx and highlighted its potential by various experiments. 
SubSpectralNets also have some limitations, including the fact that for some classes, sub-classifiers perform better than the global classifier. Also in the current model, we have to specify parameters like sub-spectrogram size and mel-bin hop-size. One way to address this could be by using the statistical analysis to choose the most appropriate parameters. In future, we plan to work on further improving the performance of this network, for example, by incorporating well-founded CNN architectures \cite{senets, densenets} or modelling the temporal information of SubSpectralNets in a more effective way. %like Squeeze-and-Excitation network \cite{senets} and Densely Connected Neural Networks \cite{densenets}. 

\newpage

\bibliographystyle{IEEEbib}
\bibliography{strings,refs}

\end{document}